\documentclass[10pt, twocolumn]{article}

\usepackage[utf8]{inputenc}
\usepackage{supertabular}
\usepackage{multirow}
\usepackage{tikz,array}
\usetikzlibrary{calc}
\usepackage{booktabs}
\usepackage{bbding}
\usepackage{tabularx}
\usepackage[utf8]{inputenc}
\usepackage[english]{babel}
\usepackage[normalem]{ulem}

\newcommand{\tikzmark}[1]{\tikz[remember picture,overlay]\coordinate (#1);}
\newcolumntype{T}[1]{@{\hspace{\tabcolsep}}c@{\hspace{\tabcolsep}\tikzmark{#1}}}
 
\newcommand{\testBox}[3]{\hspace{12pt}\llap{\rotatebox[origin=#1]{#2}{#3}}}

\usepackage{hyperref}
\usepackage{csquotes}
\hypersetup{
    colorlinks=true,
    linkcolor=black,
    filecolor=magenta,      
    urlcolor=blue,
    citecolor=blue,
    pdfpagemode=FullScreen,
}
\usepackage[capitalise]{cleveref}
\usepackage{adjustbox}

\usepackage{color, colortbl}
\definecolor{light-gray}{gray}{0.94}
\title{\bf Easier Said Than Done: The Failure of Top-Level Cybersecurity Advice for Consumer IoT Devices} 
\author{
Veerle van Harten, 
Carlos Hernández Gañán, \\
Michel van Eeten, 
and Simon Parkin \\[0.5ex]
{TU Delft, Delft, The Netherlands} \\[0.5ex]
\texttt{\{v.t.c.vanharten, c.hernandezganan, m.j.g.vaneeten, s.e.parkin\}@tudelft.nl}
}
\date{}

\begin{document}
\maketitle

\begin{abstract}
Consumer IoT devices are generally assumed to lack adequate default security, thus requiring user action. However, it may not be immediately clear to users what action to take and how. This uncertainty begs the question of what the minimum is that the user-base can reliably be asked to do as a prompt to secure their devices. To explore this question, we analyze security actions advocated at a national level and how these connect to user materials for a range of specific devices. We identify four pieces of converging advice across three nation-level initiatives. We then assess the extent to which these pieces of advice are aligned with instruction materials for 40 different IoT devices across five device classes (including device manuals and manufacturer websites). We expose a disconnect between the advice and the device materials. A stunning finding is that there is not a single assessed device to which all four top pieces of converging advice can be applied. At best, the supporting materials for 36 of the 40 devices provide sufficient information to apply just two of the four pieces of advice, typically the installation and enabling of (auto)updates. As something of a contradiction, it is necessary for a non-expert user to assess whether expert advice applies to a device. This risks additional user burden and proxy changes being made without the proposed security benefits. We propose recommendations, including that governments and researchers alike should declare their own working models of IoT devices when considering the user view.
\end{abstract}

\section{Introduction}
Currently, users of smart home devices (e.g., smart TVs, home appliances) are expected to configure and maintain these devices \cite{Reeder,Turner-2}. Many national-level initiatives have emerged, to advise consumers on how to secure their devices and to check security configurations. There has been a laudable drive to assess users' abilities to secure their smart devices (e.g., \cite{Zeng,Haney,haney2023user}). Conventionally, the usability of advice is tested by conducting a study where users' ability to apply the advice is tested. These kinds of research designs are in-depth, facilitating only a small set of advice or recommendations to be tested. Further, usability studies typically test advice which has already been qualified as applicable. In the past few years new studies have emerged that zoom out and look at the overall landscape of large corpora of security advice in the hundreds \cite{Redmiles, Reeder}, some specifically for IoT \cite{barrera2023security}, evaluating properties such as clarity and actionability. Such work can answer questions about the properties of security advice as a phenomenon (and its over-production \cite{Redmiles}), though it cannot determine if advocated actions can be applied in specific cases or environments. This can only be done by looking at the advice itself rather than involving users since that is not feasible for thousands of pieces of advice.

Between these two areas of work is a large gap which we need to address, especially for IoT security, to determine what it is that we should expect -- and encourage -- smart home device users to do, to sufficiently secure their devices. These recommendations need to function for an enormous variety of devices. Here, we approximate the user experience by looking at support materials for specific devices. This can scale to a larger set of devices than a classic user study can, while not being restricted to the advice text itself, as the large-scale studies are.

To improve our understanding of IoT security advice, with its enormous heterogeneity of devices, we propose a study design that operates in the middle ground. 
Here we evaluate the current foundation of cybersecurity guidance for IoT devices and assess whether top pieces of advice aimed at the general public align with the properties of IoT devices as described by manufacturers and third-party sources.

We select advice from three countries in the top 10 of the Network Readiness Index (NRI) of 2021 \cite{Dutta}: the United Kingdom, the United States, and the Netherlands. We investigate two research questions: (i) What is the baseline of convergent advice to users for securing consumer IoT devices?, and (ii) How does the identified baseline of convergent advice align with the content of support materials that users have access to, such as instruction manuals, videos, and organic search results?

We found divergence in what was advised to users, but also four convergent pieces of advice related to passwords and updates for IoT devices. We explore the degree to which these four pieces of convergent advice can be applied using the support materials for 40 IoT devices across five device classes (\Cref{sec:manu-meth,sec:manu-res}). Instead of purchasing each device, we utilized a scalable approach by analyzing resources that average users have at their disposal, specifically the accompanying manual, quick guide, manufacturer websites, instruction videos and organic search queries (as a means to capture features which were not mentioned in these materials). The latter includes 746 browser search results, and 626 YouTube results (212 and 76 from manufacturers, respectively), which were analyzed manually. Our contributions are as follows: 

\begin{itemize}
    \item \textbf{False applicability of not-fit-for-purpose, public-level advice.} We expose a systematic misalignment, more a break, between convergent public-level advice for users of consumer IoT devices (as provided by governments and public bodies) and security features supported in -- and prompted by -- the available device materials and online resources for a sample of 40 smart home devices. For the convergent pieces of public-level advice that were selected, we find that there is not a single device where we can apply all four pieces of advice. This means that as general advice, it is fundamentally not applicable and not fit-for-purpose, and hence is not correct advice for many devices. This is a finding that device-agnostic user studies and deep-dive, device-specific investigations have not exposed;
   
    \item \textbf{Absence of feature confirmation risks proxy changes.} Combined with the aforementioned lack of applicable top-level advice, we expose the potential for \textit{proxy changes} that offset any benefits -- it is not the case that any effort spent on security is worthwhile if it does not achieve its specific aim. We characterize a contradiction, wherein a user is required to have expertise about their device before they can understand whether security recommendations aimed at helping non-expert users apply to their devices; 
    
    \item \textbf{Scalable Internet-driven analysis of diverse consumer IoT devices.} In exploring our aim of identifying the basic steps that users of consumer IoT devices can follow, we develop and utilize a method for examining available advice in relation to device features. It scales better than user studies and expert examinations with physical devices, while still being able to evaluate the actionability of advice for individual devices, rather than being confined to the advice text itself, as the large-scale studies of security recommendations have been. 
    Given the great diversity of IoT devices, there is an urgency to find approaches that are accessible to a wider range of researchers, which are less costly and resource-exorbitant. Here, we use the accessibility of online device advice to map the existence of specific security-related device features.
\end{itemize}

We revisit our research questions in the Discussion (\Cref{sec:disc}), including recommendations such as surfacing researchers' and policymakers' own mental models of IoT devices, and a push for consistent terminology to connect high-level advice to device features. We close the paper with Conclusions (\Cref{sec:conc}).

\section{Background and Related Work}
\label{sec:back}

Here we outline the security challenges faced by smart device owners, then turn to various lines of research on security advice.

\subsection{IoT threat landscape}
The adoption of IoT devices in homes is increasing. Using the definition of Silverio-Fernández et al. \cite{Silverio}, we define an IoT device as a context-aware electronic device capable of performing autonomous computing and connecting to other devices for data exchange, in either a wired or wireless manner.

Several issues with IoT devices have been identified, such as unauthorized data collection, surveillance, and hacking \cite{Geeng, Haney, Haney-1, Ion, Zeng, Zheng}.
Once attackers realized the potential to gain unauthorized access to such systems, they started experimenting with different ways to exploit related vulnerabilities. Numerous malicious scripts, tools, and malware emerged. Malware families such as Mirai and Gafgyt \cite{Knieriem, Pham, Rodriguez, Wang} are well-known. Smart devices also have the potential to be co-opted within the home to monitor and control domestic partners, as tech-abuse \cite{lopez2019internet}. 

Numerous governments have declared baseline expectations and design principles for IoT device security. Nevertheless, these efforts are not immediate, as there are many manufacturers and product types, and a lack of reliable data on the security practices of manufacturers \cite{Rodriguez}.

\subsection{Advice as a prompt to take action}
The security of consumer devices will continue to require the involvement of end users, for the foreseeable future \cite{Cranor, Redmiles, Rostami, van2020if}. It has been observed that end users have a relatively limited understanding, and potentially erroneous or incomplete mental model, of smart devices and their associated data-processing activities \cite{Haney, Haney-1, Williams, Zheng}.

Where device users may appear to `ignore' security advice, their attitude may actually be rational when factoring in their daily activities, and to what extent manufacturers provide adequate user support \cite{Blythe, Fagan, Ion, Redmiles-2, Redmiles-4}. For example, prior to being prompted, only around half of the respondents in a study by Emami et al. \cite{Emami} expressed privacy or security concerns; this number increased to almost all respondents once prompted about these topics. This suggests that privacy and security could be latent concerns for users and must be prompted in an appropriate way, `from outside'. Many end users are interested in protecting their devices, but struggle due to lack of knowledge about security risks and protection methods \cite{Abdi, Emami}. 

According to Haney et al. \cite{Haney-1}, end users perceive they have some responsibility in securing their IoT devices, but to do so successfully requires collaboration with manufacturers and governments/regulatory organizations \cite{Nthala}. This all suggests that users' success in securing their smart devices is to some extent reliant on the efforts of other stakeholders in the wider consumer device ecosystem. The adoption of security practices can be strongly encouraged by media, family, and peers offering cybersecurity advice \cite{das2018breaking,das2014effect}, especially when unfavorable security situations are depicted with relatable people \cite{Poole, Spero, Redmiles-2}. \textcolor{black}{Nonetheless, confronted with} the overload of advice \cite{Redmiles} \textcolor{black}{and limits to user time and effort, end users may leave their devices in a less-than-secure state} \cite{Herley, Ion, Williams}. People have \textcolor{black}{limited} time to devote to \textcolor{black}{security} depending on the perceived security benefit of applying a piece of advice \cite{Fagan, Redmiles-2, Turner-2}. Other reasons for users to reject cybersecurity advice range from excessive marketing material, advice not seeming to be reliable, lack of trust in the advice source, or because the user has not yet had a negative experience \cite{Nthala, Redmiles-3}.

\subsection{Quality of consumer advice}
The quality and effects of the formulation of cybersecurity advice have been explored in several works. As Reeder et al. \cite{Reeder} point out, varying computing contexts make it challenging to derive helpful general advice. There is widespread cybersecurity advice of which most is found online, resulting in an overload of disorganized advice \cite{Redmiles}. Some pages give a sequential list of steps, assuming that the reader has a certain level of technical skill to determine how to perform them. In contrast, pages may otherwise contain so much advice that it could become overwhelming, making it difficult to know where to get started.

As a response to this problem, Ion et al. \cite{Ion} outline characteristics of sound security advice while Turner et al. \cite{Turner-2} more specifically focus on IoT devices and consider the quality of advice, how well it is written, and the extent to which it can reach end users. In line with this, \cite{Reeder} states that 'general advice' should be: Effective, Actionable, Consistent, and Concise. There, conciseness is discussed regarding the number of pieces of advice a person needs. The applicability of cybersecurity advice, however, is not taken into consideration. This issue is touched upon in only a limited sense \cite{Blythe, Haney, Haney-1}, noting a lack of available information about the security features of IoT devices, and the provision of advice on cyber hygiene by manufacturers (concluding that manufacturers do not provide a comprehensive manual or support page).

Smart home device users have also expressed dissatisfaction with the lack of support \cite{Bouwmeester}, struggling to find useful information on either the manual or support page of a device manufacturer to improve the security of their device. Regarding the applicability of advice for consumer IoT devices, \cite{Bouwmeester} demonstrates the impact of uncertainty, with end users checking for a password on their IoT device; if they cannot find it, they still do not know if it exists. As demonstrated by Reeder et al. \cite{Reeder}, good advice is not a universal truth as it is highly contextual; some pieces of advice may be effective for some people with particular computing environments, but not for others \cite{Bouwmeester, Herley, Nthala, Redmiles, Reeder, Turner}. In conclusion, there is an over-production of advice, diversity of smart home devices, and expectation for smart home users to take action; there is a gap in supporting qualifying of which advice applies to which specific devices (and in what way) and the extent to which users are supported to determine what they can do to secure their smart devices.

\section{Public-level Advice -- Methodology}
\label{sec:govt-meth}

This section addresses our first research question, `what is the baseline of convergent advice to users for securing consumer IoT devices?'. We consider governments and public bodies as acting to provide broadly applicable yet workable advice to the public. 

\subsection{Data selection} %Governmental advice
Using the top 10 of the Network Readiness Index (NRI) of 2021, we selected countries that offer specific IoT advice to citizens \cite{Dutta}. These are the United Kingdom, the United States, and the Netherlands (NL) (Dutch-language text is translated by the authors). We selected these three countries as representative of public-facing advice about IoT devices at nation-scale. Countries were also selected based on the authors having collective knowledge of the IoT and regulatory landscape of those countries. Per country, we explored the information provided for IoT devices and documented each distinct piece of cybersecurity advice. The UK and the Netherlands define IoT devices as any device that can be connected to the Internet. The US has an even broader definition, and defines an IoT device as:

\begin{quotation}{US - “Any object or device that sends and receives data automatically through the Internet. This rapidly expanding set of “things” includes tags (also known as labels or chips that automatically track objects), sensors, and devices that interact with people and share information machine to machine. \cite{CISA}.”}
\end{quotation} 
 
\subsection{Data analysis}
A 'codebook'-style thematic analysis \cite{braun2021one} was conducted to compare the pieces of advice with each other. Pieces of advice were discussed at regular codebook meetings within the author team, to identify overlaps and discuss unclear cases. An inductive approach was applied in which, per country, all pieces of advice that were given on their governmental website regarding securing IoT devices were identified. In the next stage, the gathered pieces of advice per country were considered side-by-side by the first author. Where a sufficient overlap was identified between pieces of advice of different governments, these were clustered together as one convergent piece of advice. This process was discussed in iterations with the other authors. For example, the following three pieces of advice were clustered together to formulate the convergent advice "Change default password(s) to new strong password(s)":

UK - “Consider the factory set password a placeholder. You should immediately change it the moment you start using the new device. Otherwise, anyone who previously had access to the factory settings password can access your device.” "You should make your passwords as un-guessable as possible for an outsider \cite{GetSafeOnline}."

US - “Some Internet-enabled devices are configured with default passwords to simplify setup. These default passwords are easily found online, so they don't provide any protection. Choose strong passwords to help secure your device \cite{CISA}.”

NL - “Wijzig het standaardwachtwoord en stel een sterk wachtwoord in \cite{VeiligInternetten}."(\emph{"Change the default password and set up a strong password"})

\section{Public-level Advice -- Results}
\label{sec:govt-res}

\label{sec:top-advice}
Similar to the findings of \cite{Reeder} regarding online security advice, for consumer IoT devices we found a wide spread of security advice. A total of 30 pieces of security advice were uncovered (\Cref{tbl:governmental_advice_comparison}). 

\begin{table*}[ht!]
\centering
\setlength{\tabcolsep}{2pt}
\renewcommand{\arraystretch}{1}
\begin{tabular}{rccccc}
    \multicolumn{1}{l}{\textbf{Advice}} & {\textbf{UK}} & {\textbf{US}} & {\textbf{NL}} \\
    \hline
    \rowcolor{light-gray}
    \textbf{DEFAULT CREDENTIALS} &&& \\
    %Change default to new strong password & \Checkmark & \Checkmark & \Checkmark \\
    Change default to strong password & \Checkmark & \Checkmark & \Checkmark \\
    Do not reuse passwords & \Checkmark & \Checkmark & \Checkmark  \\
    Use a password manager & \Checkmark & \Checkmark & \Checkmark \\
    Change default username & & \Checkmark & \\
    \rowcolor{light-gray}
    \textbf{ROUTER} &&& \\
    Disable UPnP & \Checkmark & \Checkmark & \Checkmark \\
    Change default password router & & \Checkmark & \Checkmark \\
    Don't use WPS & \Checkmark & \Checkmark &\\
    Disable remote management & & \Checkmark & \Checkmark\\
    Change SSID a.k.a. network name & & \Checkmark &\\
    Install a network firewall & & \Checkmark &\\
    Reduce wireless signal strength & & \Checkmark &\\
    Turn off network when not in use & & \Checkmark & \\
    Activate WPA2 & & & \Checkmark \\
    Monitor for unknown device connections & & \Checkmark & \\
    Use the router provided by the ISP & & & \Checkmark \\
    Use router intended for small businesses & \Checkmark & & \\
    \rowcolor{light-gray}
    \textbf{UPDATES} &&& \\
    Install updates & \Checkmark & \Checkmark & \Checkmark \\
    Activate automated updates & \Checkmark & \Checkmark & \Checkmark \\
    \rowcolor{light-gray}
    \textbf{NETWORK CONNECTIVITY} &&& \\
    Only connect device to internet if necessary & \Checkmark & \Checkmark & \Checkmark \\
    %Only connect your device to the internet if necessary & \Checkmark & \Checkmark & \Checkmark \\
    Use an Ethernet cable instead of Wi-Fi & & & \Checkmark \\
    \rowcolor{light-gray}
    \textbf{OTHER} &&& \\
    Create unique accounts for each user & & \Checkmark &\\
    Use multifactor authentication & \Checkmark & \Checkmark & \\
    Enable encryption features & \Checkmark & \Checkmark &  \\
    Switch off sensors if not necessary & \Checkmark & & \Checkmark\\
    %Switch the device off instead of leaving it in standby & & &\Checkmark \\
    Switch device off and not leave in standby & & &\Checkmark \\
    %Only download apps from built-in application stores & & \Checkmark & \\
    %Only download apps from application stores & & \Checkmark & \\
    Download apps from built-in app. stores & & \Checkmark & \\
    Use antivirus software & & \Checkmark & \\
    Install a firewall for IoT devices & & \Checkmark &\\
    Regularly back up your data & & \Checkmark &\\
    Remove unnecessary services and software & & \Checkmark &\\
    \hline \\[0.1cm]
\end{tabular}
\caption{Governmental Advice Comparison - The above overview shows each piece of cybersecurity advice found per country. These pieces were grouped together when various government recommendations were deemed to sufficiently overlap. This resulted in certain pieces of advice receiving more than one checkmark.}
\label{tbl:governmental_advice_comparison}
\end{table*}

Various definitions for passwords, updates, and consequences of applying these were observed. For example, within the governmental advice, there were differences between countries in whether it was declared that a device could have one, or more, `default' passwords. One explanation for this is that each country emphasizes different security aspects. The advice of the US, for example, focuses on securing the router. In contrast, the Dutch advice emphasizes the installation of updates through their campaign "Doe je updates" (\emph{"Do your updates"}). Despite these different focus points, seven pieces of convergent advice were uncovered. 

We focused on advice about securing individual IoT devices. Within this we note that although the router serves as a gatekeeper for the network, we do not consider it an IoT device. This focus resulted in putting to one side any converging advice that is not directly applicable to devices, and instead applies to the broader home network, for example, using a password manager and disabling use of the Universal Plug and Play (UPnP) protocol on the router. These pieces of advice can be considered additional measures users can take to secure their IoT devices, which play a vital role in improving the level of security of the IoT ecosystem, but are outside the scope of this research. Because we did not consider advice that obstructed actual use of the device, the advice to only connect the device to the Internet if necessary was also excluded as that defeats the purpose of using the device securely. As a result, four converging pieces of public-level advice remained\textcolor{black}{:}

\begin{enumerate}
  \item Change the default password to a new strong password.
  \item Use different passwords for different devices
  \item Install updates.
  \item Activate automated updates/set a periodic reminder in your calendar.
\end{enumerate}

\textcolor{black}{These pieces of advice were the most widely communicated top-down advice for each country and will be referred to as top pieces of advice for the remainder of this paper. An overview of the source text used for each piece of advice can be found in the Appendix. The four general pieces of convergent advice, or themes, are: to change the default password to a new strong password, to use different passwords for different devices, install updates, and activate automated updates. These top pieces of advice also align with what security experts choose as the most essential advice, as identified in existing IoT research \cite{Blythe} (and e.g., signposted in recent US government initiatives for smart home device security \cite{whitehouse2023}), but also in general cybersecurity \cite{Redmiles, Reeder}. This further implies that such advice is commonly held, and would be what reaches consumers from various community or expert channels (not just public-level advice).}

The public-level advice on the websites of the US and the UK give a summary of the cybersecurity advice they deem essential on one dedicated page \cite{CISA, GetSafeOnline}. In these pages, words are highlighted that, when clicked, redirect users to webpages that give more specific information about a particular subject. These forwarded pages contain general information, for example, on how to set a strong password aimed at computing devices and online services. 

The primary goal of the public-level advice of the selected countries is to make devices more resilient against outside threats. The threats that are mentioned are attackers, botnets, or cyber-criminals that try to break into the device with the goal to retrieve personal data, cause damage to the device, or use the device to attack other devices (in the case of a botnet attack). The UK advice also warns about risks of the misuse of personal information by the manufacturer of the device. In upcoming sections, we will further detail how these countries frame the different pieces of advice, and explore what this means for efforts to apply the device.

\subsection{Changing default password}
\label{sec:default_pwd}
When looking at the piece of advice to change the default credentials, the selected countries generally refer to changing the default password(s), as soon as possible. Only the US also recommends changing the default username. Although all the selected countries use the word \emph{default} to refer to the password they deem important to be changed, the UK uses this term interchangeably with the words `factory set password' across separate pages. The latter wording more strongly emphasizes that the advice refers to credentials set during device production. The advice of the US also suggests to consider default passwords as already public since they can be found online. In sum, users are advised to change a password on a device immediately, thereby also inferring that there is a password for the device itself. As the main objective of top pieces of advice is to make devices more resilient against outside threats, this seems to be a password that enables access over a network.

What is striking is that the advice of the UK and NL seems to assume that there is one default password per device that needs to be changed, while the US speaks of multiple passwords per device. Similar differences can be observed within the scientific literature where some works (e.g., \cite{Blythe, Knieriem}) speak in terms of \emph{the one} default password, while others (e.g., \cite{Chalhoub, Haney, Rodriguez}) more generally speak of default passwords. These interpretations suggest that each device has \emph{at the least} one set of default credentials that can be changed with little clarity as to how to be sure that a password is the one relating to network access and not another one.

\subsubsection{Setting strong password}
\label{sec:strong_pwd}
The second piece of advice is to change the default password to a strong password. All selected countries consider a password \emph{strong} if it contains upper- and lowercase letters, digits, and special characters and does not include personal information. However, the minimal length differs, varying from a minimum of 8 to 12 characters. Some advice, such as the UK and the US, discourages dictionary words and encourages random strings of letters and digits. In contrast, top  pieces of NL advice include examples of sentences with dictionary words\footnote{Parallel advice in the UK also follows this approach, as at \url{https://www.ncsc.gov.uk/blog-post/three-random-words-or-thinkrandom-0}. The latter highlights that there may also be differences in advice available \textit{within} countries}. 

\subsection{Using different passwords}
The reuse of passwords is strongly discouraged by all countries, emphasizing the importance of using `unique' passwords for devices. While the US frames this in terms of what users should not do, stating not to reuse passwords, the UK and NL emphasize actions that a user should follow: use different passwords for different devices. When changing an existing password, the US and NL directly state to set a strong and unique password, while the UK speaks of creating an 'un-guessable password for an outsider.'

\subsection{Installing updates}
As for installing updates, the security advice of all countries uses an assertive tone, urging to `apply' or `carry out' (firmware/software) updates or patches as soon as they are available. When the word `patches' is used, it is always explained that these refer to updates. In the three countries, the advice implies that the primary goal of an update is to improve the security functionality of the device, to protect against outside threats. The UK, for example, notes:
\begin{quotation}
"Firmware updates allow manufacturers to install software patches in case a security vulnerability is detected."
\end{quotation}
Taken together, this resulted in the theme `installing updates to improve the security of IoT devices'.

Though updates might include security elements that improve a device's security, this is not guaranteed as an update could potentially only -- or also -- include enhancements to features, such as modifying the interface or menus.

\subsection{Enabling auto-updates}
We generally find that users are instructed to check if a device offers the option to automatically update and manually check for updates if this is not the case.
Regarding automatic update functionality, wording includes auto-update features, automatic updates, automatic updating, and applying updates automatically. The UK and NL specifically instruct users to enable them, while the US is less direct and states to take advantage of automatic options when available. In \textcolor{black}{the absence of} auto-update capability, the US and NL governmental top pieces of advice recommend that end users periodically check for updates. 

The advice of the UK appears to assume that devices that lack an auto-update feature can be configured to notify users of new updates. Only the NL advice mentions and emphasizes a device's companion app as a way to install updates and specifies \textit{where} and \textit{how} to look for updates. Within the NL advice, it is recommended, for example, to first consult the manual to check if the device is supported with updates (and supports walking through checks for enabling automatic updates through an accompanying app). 

Where devices do not support automatic updates, it is recommended to check the settings of the companion app \textcolor{black}{for update notifications}, and if these are not there, to check the manufacturer's website. When details concerning update support are lacking, it is recommended to contact the manufacturer directly. Although the US advice covers "How do I set up automatic updates," the accompanying information similarly relies on users investigating features, such as "Turn on and confirm automatic updates" and "How you turn on automatic updates can differ depending on the software and the device" without going into more detail.

\section{Device Materials -- Methodology}
\label{sec:manu-meth}

Here we explore our second research question, `How does the identified baseline of convergent advice align with the content of instruction manuals and other related materials for the selected consumer IoT devices (as signals, or prompts) as provided -- or not -- by manufacturers?' We draw in the `top pieces of advice' from the advice provided by public-level bodies in the previous section.

\subsection{Data selection - devices and materials} 

\begin{figure*}[!ht]
    \centering
    \includegraphics[width=0.8\textwidth]{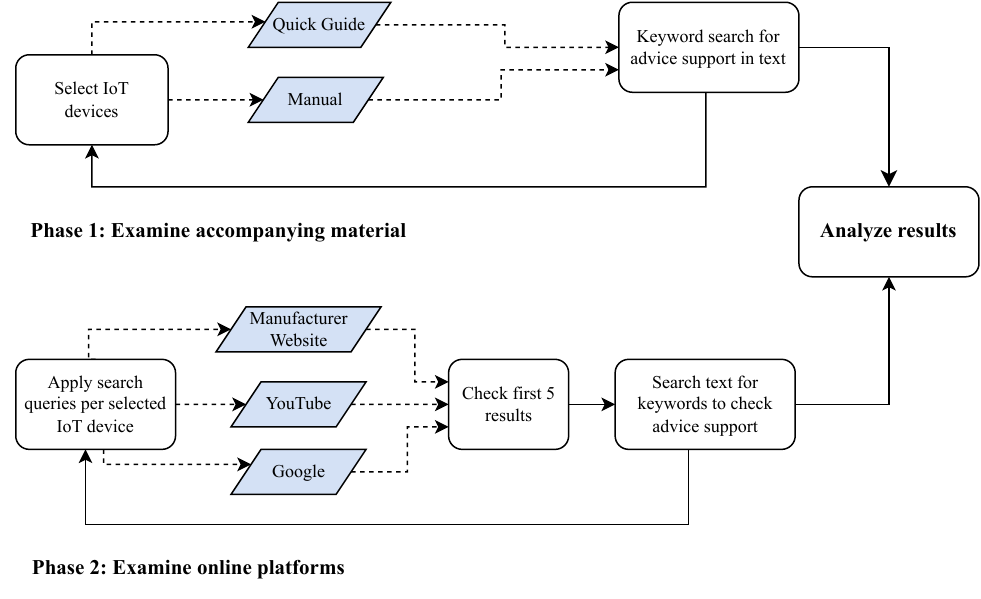}
    \caption{The examination methods utilized in this paper are shown in the above diagram. The two flow charts illustrate the analysis process for each source consulted for each IoT device. Keywords were used for the loop in the first chart and search queries for the second chart.}
    \label{fig:my_label}
\end{figure*}

In this stage, the goal is to test the presumption that the top pieces of advice align with available IoT device features. We employ a method that utilizes information sources that end users can access, in place of a physical inspection of each actual device.

We chose to rely on device documentation and Internet searches, rather than the costly and non-scalable alternative of directly purchasing each and every device, because IoT devices continue to proliferate into evermore product types and designs. As a field, we need to explore approaches that allow us to move with this proliferation. Requiring physical access to a device for analysis would severely restrict research and pose problems of generalization. 

Beside this scaling problem, approaches based on physical devices have their own drawbacks. Cognitive walkthroughs, for example, where the researcher steps through device features, have issues with ecological/external validity \cite{krol2016towards} in terms of asserting to represent the user journey \cite{lewis1993task}. Our work targets a middle-ground, or precursor step, to identify signals that qualify public-level advice as applicable to specific devices. Our approach includes third-party sources via YouTube and web search -- YouTube and top search content often responds to user needs (for missing information about devices) or relates to actual users documenting their experiences of device use (filling in gaps in support). We aim to strike a balance, by systematically exploring resources that users would have access to, bridging the top pieces of advice and device functionality.

Because the emphasis of our study is on advice to consumers, we restricted our choice of IoT devices to those utilized within a domestic setting. We selected 40 popular IoT devices across five categories, to ensure a diverse selection of commonly used IoT devices: smart entertainment, smart health, smart security, smart assistants, and smart home appliances. The release dates of the devices ranged from 2015 to 2021. Per category, we selected the devices from brands that were sold in all three countries, and listed in the top lists of frequently purchased devices per device category on the websites of popular retailer companies: Amazon, CoolBlue, and Bol.com, since these will have most of the market share. Although we examined websites available in the Netherlands, the devices we cataloged are produced by international brands, and are available in many countries. To generate an overview of known device functionality, we first consulted the manual and quick guide if available (not all devices had a 'quick guide', for example). We also consulted the website of the manufacturer and their YouTube channel (if available). 

We do not assume that all features are documented in the provided materials -- we also used a browser search for third-party search results (specifically the Google search engine) and YouTube, as manufacturer materials may not include all pertinent information about the functionality of a device. These third-party sources consisted of blogs, vlogs, forums, retailer- and news-websites, and YouTube channels. For the online sources, we used the following search queries to find more information about default passwords and how to change them, drawn from \Cref{sec:top-advice}:

\begin{enumerate}
   \item \texttt{\$DEVICE\_NAME default password}
   \item \texttt{\$DEVICE\_NAME factory set password}
   \item \texttt{\$DEVICE\_NAME changing the default password}
   \item \texttt{\$DEVICE\_NAME changing the factory set password}
\end{enumerate}
Here, \texttt{\$DEVICE\_NAME} is replaced by one of the 40 devices of which an overview can be found in \Cref{tbl:device_overview}. To find more information about the extent to which devices are supported by updates, we used the following search queries (for search (2), this included where the search engine resolved the term to `automatic'):

\begin{enumerate}
   \item \texttt{\$DEVICE\_NAME updates}
   \item \texttt{\$DEVICE\_NAME auto updates}
   \item \texttt{\$DEVICE\_NAME security updates}
\end{enumerate}

\subsection{Data collection -- accompanying material and online platforms}
Data was gathered between December 2021 and the end of March 2022. We searched for videos on YouTube, the search results in the form of webpages through Google browser search, and the website of each manufacturer, for which the first five results were checked for each search query when possible. In the case of using the search engine on the manufacturer's webpage (if there was one), rarely more than one result appeared per search query. Interestingly, on many occasions, more search results showed up for manufacturer websites through a Google browser search than when directly using the search engine on the manufacturer's website. Accounting for overlapping results, the total number of unique search results on Google was 746, and 626 on YouTube. 212 of the search results on Google and 76 on YouTube were websites and videos created by the manufacturers directly. The remainder contained websites and YouTube channels from third parties.

Within the online manuals, quick guides, written transcripts of YouTube videos, and webpages, we searched for the existence of the keywords "password" and "update." When these keywords were found in the text, it was read in order to understand the context and significance around these keywords. See \Cref{fig:my_label} for an overview of the analysis process. One author followed the same approach of codebook-style thematic analysis \cite{braun2021one} as for the resources of the top pieces of advice, involving coding and regular discussion, toward determining whether resource content supported one of the four pieces of advice. For codebook-style thematic analysis one coder is sufficient \cite{braun2021one,mcdonald2019reliability}.

\section{Device Materials -- Results}

\label{sec:manu-res}

We first checked which resources were provided by the manufacturer, \textcolor{black}{finding} that all devices in our search provide some form of documentation, such as a manual (31 devices)\textcolor{black}{,} quick guide (15), \textcolor{black}{or} both (6) (see \Cref{tbl:changing_the_default_password}). The texts were first examined using the terms \emph{default password} or \emph{factory-set password} as based on the top pieces of advice\textcolor{black}{, but in} none of the cases were the\textcolor{black}{se} terms used.

\begin{table*}[ht!]
\centering
\renewcommand{\arraystretch}{1.4}
\begin{tabular}{p{1.75in}ccccccc}
    & \testBox{rb}{315}{Quick Guide} 
    & \testBox{rb}{315}{Manual} 
    & \testBox{rb}{315}{Manufacturer Website Pages} 
    & \testBox{rb}{315}{YouTube Manufacturer Channel} 
    & \testBox{rb}{315}{YouTube Third Parties} 
    & \testBox{rb}{315}{Google WP Manufacturer Results} 
    & \testBox{rb}{315}{Google Third Parties} \\
    \hline
    \rowcolor{light-gray}
    \textbf{INFO DP APPLICABLE} &&&&&&& \\
    Changing DP to strong DP
 & 0 (0) & 0 (0)  & 0 (0)  & 0 (0) & 0 (0) & 0 (0) & 0 (0) \\
 \rowcolor{light-gray}
  \textbf{INFO PWD REUSE} &&&&&&& \\
    Discourage reuse PWDs
 & 0 (0)  & 0 (0)  & 0 (0)  & 1 (1) & 0 (0) & 0 (0) & 0 (0) \\
    \rowcolor{light-gray}
    \multicolumn{2}{l}{\textbf{INFO DP NOT APPLICABLE}} &&&&&& \\
    Changing to strong PWD
 & 0 (0) & 0 (0)  & 0 (0)  & 0 (0) & 0 (0) & 0 (0) & 0 (0) \\
    Changing DP
 & 0 (0) & 0 (0) & 0 (0) & 0 (0) & 2 (2) & 4 (5) & 7 (11) \\
    Changing/setting a PWD
 & 2 (2) & 2 (2) & 7 (7) & 2 (4) & 15 (32) & 8 (13) & 11 (12) \\
    Mentioning DP
 & 0 (0) & 0 (0) & 4 (4) & 0 (0) & 1 (1) & 1 (1) & 2 (3) \\
    Mentioning PWD
 & 0 (0) & 10 (10) & 14 (15) & 5 (6) & 20 (44) & 15 (17) & 29 (68) \\
    \rowcolor{light-gray}
    \textbf{NO INFO PWDs} &&&&&&& \\
    No info on PWD 
& 13 (13) & 19 (19) & 0 (0) & 17 (32) & 37 (203) & 28 (57) & 36 (150) \\
    Info diff. device/PNF 
& 0 (0) & 0 (0) & 40 (136) & 1 (5) & 11 (18) & 4 (5) & 12 (22) \\
    \hline
    \textbf{TOTAL UNIQUE SOURCES} 
& 15 (15) & 31 (31) & 40 (162) & 20 (48) & 40 (300) & 38 (98) & 40 (266) \\
    \hline \\[0.1cm]
%    \textbf{\% APPL SOURCES ADV 1}
%& 0\% & 0\% & 0\% & 0\% & 0\% & 0\% & 0\% \\
%    \hline
%    \textbf{\% APPL SOURCES ADV 2}
%& 0\% & 0\% & 0\% & 2\% & 0\% & 0\% & 0\% \\
%    \hline
    
\end{tabular}
\caption{Default Password (DP) Results Overview - The table provides an overview of the number of unique devices (max. 40) that are accompanied by informative material on passwords (PWD) per source category. The number of sources is listed in parentheses. The rows under ``INFO DP APPLICABLE" and ``INFO PWD REUSE" indicate the number of sources that have enough information to follow advice 1 and 2, which are to change default passwords to strong passwords and discourage password reuse. %These two rows are used to calculate the percentages in the last two rows that display the percentage chance of being able to follow each piece of advice by source category. Based on this data, it is shown that 
No sources for any device support the first piece of advice, and only one source (from the manufacturer's YouTube channel) for one device supports the second. 
The rows under ``INFO DP NOT APPLICABLE" and "NO INFO PWDs" show sources that do not have enough information to follow the advice and are included to give insight into the availability of information on (default) passwords and the frequency of lack of information to apply the pieces of advice. For example, none of the quick guides for the 40 devices mention default passwords at all.}
\label{tbl:changing_the_default_password}
\end{table*}

Interestingly, it was not uncommon to find webpages on manufacturer websites via the Google search engine (results which did not appear when using the search engine on the manufacturer website directly). A supporting overview of device coverage of advice is in Appendix, \autoref{tbl:device_overview}.

\subsection{Support for changing default password} 
The term \emph{default password} was only used for four devices on \textcolor{black}{manufacturers'} websites when using the\textcolor{black}{ir} search engine (\Cref{tbl:changing_the_default_password}, third column, `Mentioning DP'). Third-party results, such as YouTube channels and Google search results which are not the manufacturer, tended to \textcolor{black}{include} more information on default passwords, although still for the minority of devices (10)\footnote{Although \Cref{tbl:changing_the_default_password} shows 12 devices for third-party results that contain information about (changing) default passwords, two of these contained information on YouTube and Google which brings the total of devices that refer to default passwords to 10}. 

\subsubsection{Potential for many default passwords}

The phrase \emph{factory-set password} did not appear in our examination, and the results that did appear primarily included information on factory-reset\textcolor{black}{ting} device\textcolor{black}{s}. When a default password was mentioned, it did also refer to the default password on the Wi-Fi router, as was the case for e.g., Ring doorbell (Smart Security) and Bose smart speaker (Smart Assistants). Overall, Wi-Fi router\textcolor{black}{s} play\textcolor{black}{ed} a vital role within the provided resources, as the Wi-Fi password was the most mentioned password in our dataset. The primary goal of the documentation, when mentioned, was to connect the device to the Internet rather than securing the network. There are then challenges in determining if the default password for securing use of a device on the network is regarded as even being on the device itself.

\subsubsection{Qualifying a password as the default password}
For none of the devices that mention a default password, more than one default password is mentioned. This finding did not change when including the YouTube and Google results from third parties. Even with the sources mentioning a default password, it was still not possible to qualify the advice. It was, for example, not clear if it was possible to set a strong password for devices other than those that only accept numerical passwords. Furthermore, for the sources where the term `default password' was used, the instructions on how to change these were not always included, as shown in \Cref{tbl:changing_the_default_password} in categories referring to ``Info DP". In these cases, the information was mostly limited to providing information as to what device the default password is for, when someone forgets the password, or has had to reset the device. Mostly, when a default password was mentioned -- and in cases where no further instruction on how to change it was provided -- it referred to a password on the device without declaring related security functions or benefits. 

\begin{table*}[ht!]
\centering
\renewcommand{\arraystretch}{1.4}
\begin{tabular}{p{1.75in}ccccccc}
     & \testBox{rb}{315}{Quick Guide} 
    & \testBox{rb}{315}{Manual} 
    & \testBox{rb}{315}{Manufacturer Website Pages} 
    & \testBox{rb}{315}{YouTube Manufacturer Channel} 
    & \testBox{rb}{315}{YouTube Third Parties} 
    & \testBox{rb}{315}{Google WP Manufacturer Results} 
    & \testBox{rb}{315}{Google Third Parties} \\
    \hline
    \rowcolor{light-gray}
    \multicolumn{8}{l}{\textbf{INFO UPDATES APPLICABLE}} \\
    Manually installing updates 
& 1 (1) & 6 (6) & 4 (6) & 5 (5) & 16 (31) & 8 (9) & 18 (21) \\
    Enabling/forced auto-updates 
& 0 (0)  & 6 (6)  & 5 (6)  & 1 (1) & 9 (16) & 12 (17) & 9 (28) \\
    \rowcolor{light-gray}
    \multicolumn{8}{l}{\textbf{INFO UPDATES NOT APPLICABLE}}  \\
    Info about auto-updates 
& 0 (0) & 1 (1) & 2 (2) & 0 (0) & 3 (4) & 8 (13) & 16 (20) \\
    Contains info about updates 
& 1 (1) & 1 (1) & 6 (7) & 1 (1) & 21 (45) & 13 (26) & 34 (83) \\
    \rowcolor{light-gray}
    \textbf{NO INFO UPDATES} &&&&&&& \\
    No info about updates 
& 13 (13) & 17 (17) & 0 (0) & 18 (31) & 38 (207) & 26 (48) & 30 (109) \\
    Info diff. device/PNF  
& 0 (0) & 0 (0) & 28 (33) & 1 (1) & 11 (26) & 4 (7) & 11 (26) \\
    \hline
    \textbf{TOTAL UNIQUE SOURCES} 
& 15 (15) & 31 (31) & 40 (54) & 21 (39) & 40 (329) & 38 (120) & 40 (287) \\
    \hline \\[0.1cm]
%    \textbf{\% APPL SOURCES ADV 3} 
%& 7\% & 19\% & 11\% & 13\% & 9\% & 8\% & 7\% \\
%    \hline
%    \textbf{\% APPL SOURCES ADV 4} 
%& 0\% & 19\% & 11\% & 3\% & 5\% & 14\% & 10\% \\
%    \hline

\end{tabular}
\caption{(Auto-)Update Results Overview - The table provides an overview of the number of unique devices (max. 40) that are accompanied by informative material or not on (auto) updates per source category. The number of sources is listed between parentheses. The rows under ``INFO UPDATES APPLICABLE" indicate the number of sources that have enough information to follow advice 3 and 4, which are to manually install updates and the enabling of auto-updates. It's worth noting that all devices have at least some sources that mention updates, but that the majority of sources do not provide information on how to install or enable (auto)updates. The rows under ``INFO UPDATES NOT APPLICABLE" and ``NO INFO UPDATES" show sources that do not have enough information to follow the advice and give insight into the availability of information on (auto) updates and the frequency of lack of information to apply the pieces of advice.}
\label{tbl:auto_update_support}
\end{table*}

\subsubsection{Having multiple default passwords for securing a device}
Our analysis shows that a great variety of passwords can be changed, which are mostly not referred to as the 'default' password. 

An encouraging example of manufacturers recognizing this challenge is the Reolink doorbell (Smart Security), which dedicates a webpage to describing the distinctions between the Password for the Reolink App, the Reolink Client, and the Reolink Cameras (Smart Security). The password(s) that can be changed on devices may still represent one or several default passwords even while the phrase "default password" is missing. In these cases, any password on the device could or could not be a default password, making it difficult to determine whether it is possible to change the default password as advised, as shown by the Reolink example. On top of that, even when the wording 'default password' is used, it does not necessarily provide security benefits against outside threats, like in the cases where default passwords refer to a 4-digit parental code (as for two smart TVs, the Samsung UE49MU8000 and the LG UHD TV 43UP80 (Smart Entertainment). 

As another example, the default password of two smart TVs only allowed for setting a 4-digit code. Even if it had been possible to set a strong password in these instances, it still would not protect end users from outside threats that were described in the top pieces of advice sources. This is because these kinds of passwords, as a class of parental controls, only protect against the change of TV settings by unauthorized members within the home, and are not associated with the root control of the device (that could potentially be exploited over the network).

\subsubsection{A difference between strong and stronger passwords}
If the existing password did not meet the recommend\textcolor{black}{ations} for a strong password, the advice to set a strong password can provide security benefits, but only if the device supports these requirements. However, as shown in \Cref{tbl:changing_the_default_password}, we did not find a single resource \textcolor{black}{confirming} that it is possible to set a password that complied with the requirements of the top pieces of advice for setting a strong password. \textcolor{black}{S}ources providing information \textcolor{black}{on} setting or changing a password\textcolor{black}{, not limited to digits only}, indicate a requirement of using a minimum number of characters and allowing for letters, and digits to be set.

For eight devices it was mentioned \emph{how} to change the default password, see row ``Changing DP" in \cref{tbl:changing_the_default_password}. However, this did not mean it could be changed to a password that contained upper-and lowercase letters, digits, and special characters (as often advised, \autoref{sec:strong_pwd}). 

For the NVIDIA media player (Smart Entertainment), an explanation of how to change the default password was found, but it was not made clear if it was possible to apply the criteria of the top pieces of advice for setting a strong password. This marks the difference between being able to make a password sufficiently strong, or as strong as the device allows it to be.

In some cases, it is mentioned to use upper- and lower-case letters, but never in combination with the use of special characters. The latter is even actively discouraged when creating a Govee smart light (Smart Entertainment) user account. They note, ``Passwords should be 8–20 characters using both letters and numbers only. Do not include special characters or symbols." 

The Anova smart oven (Smart Home Appliances) and Withings smart scale (Smart Health) \textcolor{black}{require} lowering the router's security as these can only connect to it if the router's password \textcolor{black}{lacks} special characters. It can be concerning that all devices seem to allow, and some even force, setting a password that does not meet the requirements of the top pieces of advice. This permission for the use of weak passwords is, unfortunately, a finding that is not uncommon (see also \cite{Ipsos}). 

\subsection{Support for using unique passwords}
Regarding the use of unique passwords, only in the case of the Reolink doorbell (Smart Security) is there one YouTube video from the official channel of Reolink where the reuse of passwords is actively discouraged, which shows that the theme where passwords are understood as a way to protect against unauthorized access is shared to some extent by manufacturers, see \Cref{tbl:changing_the_default_password}.

\subsubsection{Attempt to apply password advice scenarios}
Taken together, the range of advice we observed in relation to changing the default password to a new strong password could, in practice, translate to the following scenarios:

\begin{enumerate}
   \item The end user changes a default password with security benefits against outside threats. Of the 6 manufacturers where a default password was mentioned in their material, 3 could potentially offer security benefits to outside threats, however it was not clear if these could be changed, let alone to a strong password.
   \item The user changes a password called the \emph{default password} by the manufacturer that does not add any security benefits regarding online threats, which was the case for 2 manufacturers (`default password' was a parental PIN code).
   \item There is no information about a default password, which was the case for 30 devices, so the end user does nothing.
   \item There is no information about a default password, so the end user changes a password that they can find for the device, that does not confer security benefits to outside threats. This would be possible, especially if the device has multiple accounts associated with it which each have a password or PIN, where potentially only one confers the security benefits described in the top pieces of advice. 
   \item There is no information about a default password, as was the case for 75\% of the analyzed sources; the end user changes a password that does have security benefits. 
   \item There is no information about a default password, so the user changes a password for another device, such as changing the Wi-Fi password on the router. At best, this could inadvertently result in a security benefit.
   \item There is no information about a default password, but the end user searches for one anyway under the assumption that there is one. This may leave the user in an ambiguous state, either satisfied that they have checked, or concerned that they have not been able to secure the device.
\end{enumerate}

Scenario 2 above could potentially lead to end users feeling they have applied the advice while expecting additional security benefits, even though this is not the case, leaving them feeling more secure than they are. Alternatively, in Scenarios 3-6, where end users cannot be sure that the advice applies to their device because the information is missing, it is much less assured that the intended security benefit will happen, appearing more like a `folk' security behaviour \cite{wash2010folk}. The opposite is also possible, where end users can feel less secure when not able to apply advocated advice. There is existing evidence of the potential for users to be unsure of the security features and benefits provided by their IoT devices \cite{Zeng}.

\begin{table*}[ht!]
\centering
\renewcommand{\arraystretch}{1.4}
\begin{tabular}{lccccccc}
    & \testBox{rb}{315}{Quick Guide} 
    & \testBox{rb}{315}{Manual} 
    & \testBox{rb}{315}{Manufacturer Website Pages} 
    & \testBox{rb}{315}{YouTube Manufacturer Channel} 
    & \testBox{rb}{315}{YouTube Third Parties} 
    & \testBox{rb}{315}{Google WP Manufacturer Results} 
    & \testBox{rb}{315}{Google Third Parties}
 \\
    \hline
    \rowcolor{light-gray}
    Smart Entertainment
 & 0(0) & 4(4) & 3(2) & 2(2) & 16(5) & 4(3) & 6(5) \\
    Smart Health
 & 1(1) & 3(3) & 1(1) & 1(1) & 7(2) & 3(3) & 6(4) \\
    \rowcolor{light-gray}
    Smart Security
 & 0(0) & 2 (2) & 2(2) & 1(1) & 9(4) & 6(6) & 7(4) \\
    Smart Assistants
 & 0(0) & 1(1) & 6(4) & 1(1) & 14(7) & 11(6) & 27(7) \\
    \rowcolor{light-gray}
    Smart Home Appliances
 & 0(0) & 2(2) & 0(0) & 1(1) & 1(1) & 2(1) & 3(3) \\
    \hline
    \textbf{TOTAL UNIQUE SOURCES} 
 & 1(1) & 12(12) & 12(9) & 6(6) & 47(19) & 26(19) & 49(23) \\
    \hline  \\[0.1cm]
    
\end{tabular}
\caption{Applicable (Auto-)Update support per IoT device category - The above table represents the distribution of sources per device category that support the applicability of the advice to install or enable auto-updates. The number of unique devices (max. 40) to which these sources apply is listed between parentheses.}
\label{tbl:updates_per_IoT_device}
\end{table*}

\subsection{Support for installing updates} 
When checking the quick guides of devices, only the Garmin smart scale (Smart Health) and Wink smart home hub (Smart Assistants) \textcolor{black}{mentioned} update support. The first show\textcolor{black}{ed} what the icon looks like when the device is installing an update and when an update was successful or failed\textcolor{black}{, while the other briefly} describe\textcolor{black}{d} how to update the app. Although still low, consulting the manual offered a higher chance of finding information about the provision of updates, as this was provided for 15 devices. Including \textcolor{black}{manufacturers'} webpages and YouTube channels resulted in a stark increase to 34 devices that offer update-support, which grew further to all devices when including non-manufacturer webpages and YouTube channels. \textcolor{black}{However, this information wasn't always easy to find, as seen in} \Cref{tbl:auto_update_support,tbl:auto_update_support_score}. Similar to \Cref{tbl:changing_the_default_password}, most device materials in \Cref{tbl:auto_update_support} were not aligned with the top pieces of advice demonstrat\textcolor{black}{ing} that features exist\textcolor{black}{ed} for devices but were not adequately documented.

There were significant differences between device categories regarding the amount of sufficient information provided. It was, for example, hard to find information on updates for smart home appliances if compared to smart assistants; see \Cref{tbl:updates_per_IoT_device}. In cases where \textcolor{black}{update} information was found, it was not always sufficient to apply the advice to manually install updates. In other cases, information on manually install\textcolor{black}{ing} updates was missing or not intended for the user to do\textcolor{black}{, as with} the LG smart refrigerator (Smart Appliances), where a YouTube video showed that only LG Authorized Service \textcolor{black}{could} update the software.

The purpose of an update is broader than is the case for a password and primarily \textcolor{black}{serve to repair, improve, or add functionalities}. The general assumption in the advice is that security issues are fixed through updates. We then also checked for the mention of security updates. Similar to \cite{Blythe}, 11\% of the results \textcolor{black}{mentioned security updates, usually limited to disclosing a patch or update applied.}

\subsection{Support for enabling auto-updates}
For devices where we found information on security updates/security patches (as a subset of our sample, as in \autoref{tbl:auto_update_support_score}), information about auto-updates was also provided in half of these cases. Most manufacturers encourage enabling automatic updates with the promise that they will improve the device's performance. Only for the Samsung smartwatch (Smart Health), it was clearly stated that automatic update support is not provided (which would help to reduce uncertainty, a noted use concern elsewhere \cite{Haney-2}). The promise to make the device more secure is also mentioned but less frequently. It shows that manufacturers, leading to the development of the fourth theme, mostly present updates as a way to improve the functionality and usability of IoT devices. This mirrors findings elsewhere \cite{Haney}, wherein users appear not to relate device updates to security functionality.

\begin{table*}[ht!]
\centering
\renewcommand{\arraystretch}{1.4}
\begin{tabular}{lccccccc}
    & \testBox{rb}{315}{Manual-updates only} 
    & \testBox{rb}{315}{Auto-updates only} 
    & \testBox{rb}{315}{Manual \& auto-updates}
    & \testBox{rb}{315}{Not provided}
 \\
    \hline
    \rowcolor{light-gray}
    1. All sources make mention updates. 
& 0 & 0 & 4 & 0 \\
    2. More than half of the sources make mention of updates.
& 2 & 3 & 10 & 0 \\
    \rowcolor{light-gray}
    3. Half of the sources make mention of updates.
& 5 & 2 & 5 & 0 \\
    4. Less than half of the sources make mention of updates.
& 1 & 0 & 1 & 0 \\
    \rowcolor{light-gray}
    5. Only one of the sources makes mention of updates.
& 5 & 1 & 0 & 1 \\
    \hline
    \textbf{TOTAL DEVICES} 
& 13 & 6 & 20 & 1 \\
    \hline  \\[0.1cm]
    
\end{tabular}
\caption{(Auto-)Update Effort Score - For each source it is checked if and what information was provided regarding update support using the manufacturers' and non-manufacturers' sources for each device. All devices contain at least one source that provides information regarding update support; however, for one device, the kind of update support, whether manual, automatic, or both, was not provided.}
\label{tbl:auto_update_support_score}
\end{table*}
The top pieces of advice \textcolor{black}{imply} that users need to enable automatic updates manually. However, 12 of the 16 devices seem to update themselves by default when connected to the internet (the remaining four mention that it can be switched on). This automation translates the advice from enabling updates to declaring no need for users to take action (which could be just as useful to know). This minimization of end user involvement connects to the ethical considerations described by van Steen \cite{Steen}, as a decision is made by the manufacturer that is not always (clearly) communicated to users and restricts freedom to make their own choices. 

For devices where it is clear that updates are separate for the device and companion app, this does not imply they both support automatic updates by default. The Google smart doorbell (Smart Security), for example, updates automatically by default, whereas the accompanying app does not. 

\section{Discussion}
\label{sec:disc}

Returning to our first research question, we identified four themes of converging public-level advice for consumer IoT devices: change default passwords; enable automated updates; and install (manual) updates. Within our limited sample, the terminology was inconsistent across advice sources, but with overlap on some points.

When considering our second question and the content of manufacturer materials, the current way that pieces of general advice are formulated may seem reasonable at first glance, but does not connect with the features indicated by the manufacturer-provided information for devices. This requires the user to infer there is a connection, where relying on some existing knowledge of terms and what they mean has its shortcomings \cite{Reeder}. For example, for some devices not every credential could be used to log into the device via the network. This misses opportunities to leverage \textit{prompts} \cite{fogg2019tiny} during device configuration to increase security.

Update features existed for all examined devices (as determined in our broader search of third-party sources), but \textcolor{black}{were} barely documented by manufacturers, with an 11\% likelihood of discovering this information inside our dataset when only focusing on \textcolor{black}{manufacturer-provided material}. See \Cref{tbl:auto_update_support}. In device materials, there is then a lack of explicit declaration of the existence (or not) of security features. 

Combining our two research questions, informing our aim of identifying \textit{basic non-trivial advice} for consumer IoT users to follow, the public-level advice that we examined mostly could not be directly related to our set of selected (popular) devices, meaning that it is not targeted, correct advice for increasing the security of the device -- features are often not there, not confirmed, or not mentioned. Based on the sources analyzed, none of the devices seem to support all four top pieces of advice at once. Device materials rarely mentioned passwords, then relying on software/apps to provide just-in-time prompts and signaling, for both the purpose and expectations for password security, in the absence of explanation elsewhere.

Compounding these issues, terminology \textcolor{black}{discrepancies also existed between} the top pieces of advice for passwords and device materials. For example, where `default' passwords were mentioned in device materials, it could also refer to the router/Wi-Fi password. \textcolor{black}{Consequently}, in many cases there is no direct route from public-level advice to device materials to user actions. 

\subsection{Acknowledging the mismatch}

\textcolor{black}{Current a}dvice bodies offer generic pieces of cybersecurity advice that, \textcolor{black}{despite} the IoT environment \textcolor{black}{being} highly diverse, convey a sense that IoT devices are alike. It is not that the advice loses detail because \textcolor{black}{it} is generalized \cite{Reeder}, but because it is \textit{selectively applicable}; \textcolor{black}{by} not considering the diversity of IoT devices, there is a lack of assurance and signaling whether advice is \textit{relevant to specific devices} before it is applied; this is left to the user. 

The inconsistent applicability of IoT advice to devices sits alongside it appearing to be applicable where it is not. This results in an intervention that works for devices where it applies, but has a different effect -- or potential unintended side-effects \cite{Osman} -- for devices which the advice does not match to, in terms of whether those features exist for those devices. The reliance on users to qualify IoT advice also brings unintended side effects \cite{chua2019identifying}. Government advice is not `wrong' for \textcolor{black}{compliant} devices, but our analysis suggests such devices are, by far, in the minority. \textit{proxy changes} \cite{Osman} could mean that `some' security was improved, e.g., a child-protection PIN as we saw for some devices, but not security against network-based attacks.

The burden appears to be on the user to determine -- to almost know in advance, despite being assumed to be non-experts -- which specific devices advice applies to. Any users of the 40 devices in our study would need to somehow know enough to decide -- while lacking confirmatory information -- that at least two and often three of the top pieces of advice do not apply to their device (as we found in our sample). For example, there may be multiple passwords or not, of which one or more may be a `default', relating to network accessibility or not, with a capacity to be of variable attainable strength. Recent research already highlights users being in a gulf between assuming features do not exist and not being aware of them (as with smart device updates \cite{Haney,haney2023user}), with advice not being specific enough to find the relevant feature. Transparency in how features work can inform such user decisions around smart device security features \cite{ponticello2021exploring}. For instance, users whose devices rely on manual updates may not know that they do \cite{haney2022user}, assuming instead that update installation is automated. Prior work has also evidenced that users who assume a device has a password may explore a range of sources and still not be able to determine if a password feature exists \cite{Bouwmeester}.

There is a mismatch between current, diverse devices and a future-looking regulatory regime (e.g., EU RED \cite{European}, and also e.g., the UK Code of Practice for Consumer IoT \cite{ukCode}). Further, there is a lack of consideration for the impact on user behaviors while generalized IoT advice and device features remain out of alignment.

\subsection{Limitations}
\label{sec:limit}

Our choice to analyze support materials, rather than the physical devices themselves, has certain pros and cons. A downside is that we did not verify firsthand the features of the selected devices, or infer their features through companion apps. Some features may be brought to the user's attention through just-in-time notifications on the companion app, for instance; however, this would only emphasize the reliance on the user to realize the relevance of any security prompts that exist, for lack of signaling from support sources. There is a difference between a device having particular features or not, and whether this is mentioned clearly in any associated instruction materials. We chose to rely on device documentation and Internet searches, rather than the costly and non-scalable alternative of directly purchasing each and every device. Actual devices could be assessed by the researcher, but related approaches such as cognitive walkthroughs have their own issues with ecological validity \cite{krol2016towards,lewis1993task} and consistency in mapping device features.

\subsection{Recommendations}
Based on our findings, we arrive at the following recommendations:

\begin{itemize}
    \item \textbf{Researchers / Policymakers: surface your own mental models of device features as well as those of users.} Significant progress has been made in understanding users' perceptions of device functionality, for instance, in exploring the IoT threat models perceived by users \cite{Zeng}. In our examination of public-level advice and device materials, we found a disparity that top pieces of advice appeared to assume that features exist, which we then did not find evidence of for all of our selected devices (as in \autoref{tbl:changing_the_default_password} and \autoref{tbl:auto_update_support}). This suggests a generic model of a consumer IoT device, that has mostly escaped scrutiny, as the kind of `common standard' that, e.g., Blythe et al. \cite{Blythe} anticipate or expect.  It is important to document assumptions about the functionality of devices that advice is being provided for and that users interact with, as part of data-gathering. For instance, a user may struggle to find the default password on a device not because of a lack of security knowledge, but because the feature is not provided for their device(s) (as we found for many devices, see \autoref{tbl:changing_the_default_password}). Steps in this direction are being made, e.g., documenting participants' devices \cite{Zeng,Zheng}. We must connect devices to specific responses from users (and researcher assumptions of device capabilities), to understand if users are struggling to use a feature because of its difficulty, or its absence.
    \item \textbf{Policymakers: match advice to groups of devices.} It may be possible to identify distinct classes or groups of devices which are at the very least \textit{more likely} to have the features that advice refers to. This would act as a shortcut that removes the need for users to determine if their device would benefit from the advice. We found, as in \autoref{tbl:updates_per_IoT_device}, that e.g., information on (automatic) update support was documented more for smart assistants than for smart home appliances. 
    \item \textbf{Manufacturers: declare the existence of security features.} We posit that if a security feature is included in a device, it should be made known that it exists and how it works. As demonstrated by the results in \autoref{tbl:changing_the_default_password}, when a default password was mentioned, information on how to change it was lacking. Also, instructions for installing updates or setting auto-updates was not always provided (\Cref{tbl:auto_update_support,tbl:auto_update_support_score}). The key is not to assume that users will know about the existence of features (irrespective of knowing how to use them). This relates to encouraging manufacturers to be more open about device functionality \cite{Haney-2}, but ideally in a way that also relates conveying security to customers (so they can both use and trust the device). Act first with an assumption that the feature is new to the user and that concrete steps are needed \cite{Prange}, rather than it being familiar \cite{fogg2010behavior,Haney-2}.
    \item \textbf{Policymakers / Manufacturers: consistent, approachable terminology.} We found, for example, that `default password' had different meanings and functionalities depending on the type of device (see \autoref{sec:default_pwd}), and that the advice offered by the studied countries differed in the number of credentials a device may use. It would benefit the disparate activities of researchers, policymakers, and manufacturers to have a narrower body of terminology that links advice to device materials and features, as a two-stage process of advising users. Prior work identifies this as a manufacturer responsibility \cite{fagan2020foundational}, where there is also a role for policymakers in ensuring the applicability of advice terminology to device features.% when advice is intended to point to concrete user action.
\end{itemize}

\section{Conclusion}
\label{sec:conc}

We identified four pieces of convergent advice to users across three representative countries (UK, US, and the Netherlands). No device was found to which all four pieces of converging governmental advice could be applied, suggesting that the advice was developed with an exemplar device in mind which at best is not like the majority of devices on the consumer market.
These findings question the value of high-level advice campaigns by governments or industry. Broadly speaking, future research will address these issues by exploring the degree to which users can be provided with more specific cybersecurity advice for their IoT device while also considering the requirements this places on other stakeholders such as policymakers maintaining advice and manufacturers of specific devices.

\bibliographystyle{IEEEtran}
\bibliography{references}

\newpage
\cleardoublepage
\newpage

\appendix

\section{The 4 top pieces of advice}

\subsection*{Advice 1: Changing the default password}

\begin{enumerate}
    \item UK - “Consider the factory set password a placeholder. You should immediately change it the moment you start using the new device. Otherwise, anyone who previously had access to the factory settings password can access your device.”
    \item US - “Some Internet-enabled devices are configured with default passwords to simplify setup. These default passwords are easily found online, so they don't provide any protection. Choose strong passwords to help secure your device.”
    \item NL - “Wijzig het standaardwachtwoord en stel een sterk wachtwoord in."("Change the default password and set up a strong password")
\end{enumerate}

\subsection*{Advice 2: Use different passwords for different devices}
\begin{enumerate}
    \item UK - "Use different passwords for different devices. Don’t reuse the same, or even similar, passwords across the many cloud-based gadgets you own. If one device is compromised, it could put all other devices at risk. Getting rid of careless habits like this is essential for keeping your home network secure."
    \item US - "Reusing a password, even a strong one, endangers your accounts just as much as using a weak password. If attackers guess your password, they would have access to your other accounts with the same password." "Use the following techniques to develop unique passwords for each of your accounts: Use different passwords on different systems and accounts."
    \item NL - "Gebruik verschillende wachtwoorden voor verschillende accounts."\\
    ("Use different passwords for different accounts")
\end{enumerate}

\subsection*{Advice 3: Install updates}

\begin{enumerate}
    \item UK - "Keep device software up to date." "Firmware updates allow manufacturers to install software patches in case a security vulnerability is detected."
    \item US - "Ensure you have up-to-date software." "When manufacturers become aware of vulnerabilities in their products, they often issue patches to fix the problem. Patches are software updates that fix a particular issue or vulnerability within your device’s software." "Make sure to apply relevant patches as soon as possible to protect your devices."
    \item NL - "Krijg je een melding om een update te doen? Doe het direct. Zie je in het scherm van je smartphone, tablet of laptop been update-notificatie van je slimme apparaat? Voer de update direct uit. Zo blijft je slimme apparaat veilig. En weet je zeker dat je up-to-date bent."\\
    ("Do you get a notification to do an update? Do it immediately. Do you see an update notification from your smart device on the screen of your smartphone, tablet, or laptop? Run the update immediately. This keeps your smart device safe. And you can be sure that you are up-to-date.")
\end{enumerate}

\subsection*{Advice 4: Activate automated updates/set a periodic reminder in your calendar}
\begin{enumerate}
    \item UK - "Some devices, like your smartphone, can auto-update security fixes. On the other hand, the router, the smart light bulbs, or the smart fridge might not. If auto update features are available, enable them on all your devices. If not, set up the device so you get alerts for newly available firmware updates. Then you can immediately update your devices on time."
    \item US - "Some software will automatically check for updates, and many vendors offer users the option to receive updates automatically. If automatic options are available, the Cybersecurity and Infrastructure Security Agency (CISA) recommends that you take advantage of them. If they are not available, periodically check your vendor’s websites for updates."
    "What is the difference between manual and automatic updates? Users can install updates manually or elect for their software programs to update automatically. Manual updates require the user or administrator to visit the vendor’s website to download and install software files. Automatic updates require user or administrator consent when installing or configuring the software. Once you consent to automatic updates, software updates are “pushed” (or installed) to your system automatically."
    \item NL - "Hoe update ik mijn slimme apparaten? Stap 1. Open de app van je slimme apparaat. Stap 2. Kijk bij ‘Instellingen’. Stap 3. Klik daar op ‘Updaten’. Stap 4. Stel ‘Automatisch updaten’ in. Soms is automatisch updaten niet mogelijk. Zet dan een herinnering in je agenda om elke 1e dag van het nieuwe kwartaal je updates te chicken. Krijg je een melding om een update te doen? Doe het direct. Heb je geen app van je slimme apparaat? Ga dan naar de website van de fabrikant en zoek op 'update'."\\ 
    ("How do I update my smart devices? Step 1. Open the app of your smart device. Step 2. Look at 'Settings'. Step 3. Click on 'Update' there. Step 4. Set up 'Automatic update'. Sometimes automatic updating is not possible. In that case you can put a reminder in your calendar to check your updates every 1st day of the new quarter. Do you get a notification to do an update? Do it right away. Don't have an app for your smart device? Go to the manufacturer's website and search for 'update.'")
\end{enumerate}

\newpage
\begin{table*}[ht!]
\section{IoT devices comparison}
\centering
\renewcommand{\arraystretch}{1}
\scriptsize
\begin{tabular}{lccccc}
    & \testBox{rb}{315}{Advice I} 
    & \testBox{rb}{315}{Advice II} 
    & \testBox{rb}{315}{Advice III} 
    & \testBox{rb}{315}{Advice IV}
 \\
            \hline
            \rowcolor{light-gray}
            \multicolumn{1}{r}{\textbf{SMART ENTERTAINMENT}} &&&& \\
            \multicolumn{1}{r}{Samsung UE49MU8000} & & & \Checkmark & \Checkmark \\
            \multicolumn{1}{r}{LG UHD TV 43UP80} & & & \Checkmark & \Checkmark \\
            \multicolumn{1}{r}{Philips PUS8506} & & & \Checkmark & \Checkmark \\
            \multicolumn{1}{r}{Hue Amaze EMEA} & & & \Checkmark & \Checkmark \\
            \multicolumn{1}{r}{Lumiman Smart Wifi Light Bulbs} & & & \Checkmark &\\
            \multicolumn{1}{r}{Govee LED Light Bulb} & & & \Checkmark &\\
            \multicolumn{1}{r}{NEBULA Anker Capsule, Smart Wi-Fi Mini Projector} & & & \Checkmark &\\
            \multicolumn{1}{r}{NVIDIA SHIELD Android TV Pro 4K HDR Streaming Media Player} & & & \Checkmark & \Checkmark\\
            
            \rowcolor{light-gray}
            \multicolumn{1}{r}{ \textbf{SMART HEALTH}} &&&& \\
            \multicolumn{1}{r}{Withings Body Cardio} & & & \Checkmark &\\
            \multicolumn{1}{r}{Garmin Index S2} & & & \Checkmark &\\
            \multicolumn{1}{r}{Arboleaf Smart Scale} & & & \Checkmark &\\
            \multicolumn{1}{r}{Xiaomi Mi Air Purifier 3H} & & & \Checkmark &\\
            \multicolumn{1}{r}{Fossil Monroe Hybrid HR Watch FTW7039} & & & &\\
            \multicolumn{1}{r}{Samsung Galaxy Watch 4 Classic} & & & \Checkmark &\\
            \multicolumn{1}{r}{Peloton Bike+} & & & \Checkmark &\\
            \multicolumn{1}{r}{NordicTrack Commercial 2950 Treadmill} & & & \Checkmark &\\
            
            \rowcolor{light-gray}
            \multicolumn{1}{r}{\textbf{SMART SECURITY}} & & & &\\
            \multicolumn{1}{r}{Reolink Argus 3 Pro} & & \Checkmark & \Checkmark &\\
            \multicolumn{1}{r}{Google Nest Cam IQ Outdoor} & & & & \Checkmark \\
            \multicolumn{1}{r}{V380 Pro} & & & \Checkmark & \\
            \multicolumn{1}{r}{Ring Pro 2} & & & \Checkmark & \Checkmark \\
            \multicolumn{1}{r}{UniFi G4 Doorbell} & & & \Checkmark &\\
            \multicolumn{1}{r}{Eufy 2K Battery Doorbell} & & & \Checkmark & \Checkmark \\
            \multicolumn{1}{r}{Fibaro Smoke Sensor} & & & \Checkmark &\\
            \multicolumn{1}{r}{Bold Smart Lock SX-33} & & & \Checkmark &\\
            
            \rowcolor{light-gray}
            \multicolumn{1}{r}{\textbf{SMART ASSISTANTS}} &&&& \\
            \multicolumn{1}{r}{Echo Studio} & & & & \Checkmark\\
            \multicolumn{1}{r}{Nest Audio} & & & & \Checkmark \\
            \multicolumn{1}{r}{Sonos One Smart Speaker} & & & \Checkmark & \Checkmark \\
            \multicolumn{1}{r}{Apple HomePod mini} & & & \Checkmark & \Checkmark \\
            \multicolumn{1}{r}{Athom Homey Pro} & & & \Checkmark & \Checkmark \\
            \multicolumn{1}{r}{JBL Link Portable} & & & \Checkmark & \\
            \multicolumn{1}{r}{Wink Hub 2} & & & \Checkmark & \Checkmark \\
            \multicolumn{1}{r}{Bose Home Speaker 500} & & & \Checkmark & \Checkmark \\
            
            \rowcolor{light-gray}
            \multicolumn{1}{r}{\textbf{SMART HOME APPLIANCES}} &&&& \\
            \multicolumn{1}{r}{iRobot Roomba J7+} & & & & \Checkmark \\
            \multicolumn{1}{r}{Xiaomi Viomi S9} & & & \Checkmark &\\
            \multicolumn{1}{r}{Miele DG 7140} & & & &\\
            \multicolumn{1}{r}{LG GC-X247CSAV Refrigerator} & & & \Checkmark &\\
            \multicolumn{1}{r}{Anova Precision Oven} & & & \Checkmark &\\
            \multicolumn{1}{r}{Instant Pot Smart WiFi} & & & &\\
            \multicolumn{1}{r}{Hamilton Beach Smart Coffee Maker} & & & &\\
            \multicolumn{1}{r}{Philips Sonicare Prestige 9900 HX9992/12} & & & \Checkmark &\\
            \hline
\end{tabular}
\vspace*{10pt}
\caption{IoT comparison table - Here we present an overview of the 40 selected IoT devices and the extent that each piece of advice (Advice I: Changing the default password, Advice II: Use different passwords for different devices, Advice III: Install updates, IV: Activate automated updates/set a periodic reminder in your calendar) could be applied.}
\label{tbl:device_overview}
\end{table*}
\end{document}